\documentclass[12pt]{iopart}

\usepackage{iopams}  
\usepackage{graphicx}
\usepackage{epsfig}
\usepackage{delarray}
\usepackage{bm}

\newcommand{\ket}[1]{| \, #1 \rangle}
\newcommand{\bra}[1]{ \langle #1 \,  |}

\begin{document}
\title{Local expansion of photonic W state using a polarization dependent beamsplitter}

\author{Toshiyuki Tashima$^{1,2}$,\c{ S}ahin Kaya \"Ozdemir$^{1,2,3}$,\\ Takashi Yamamoto$^{1,2}$, Masato Koashi$^{1,2}$, and Nobuyuki Imoto$^{1,2}$}
\address{$^1$ Division of Materials Physics, Department of Materials Engineering Science,
Graduate School of Engineering Science, Osaka University,
Toyonaka, Osaka 560-8531, Japan}
\address{$^2$ CREST Photonic Quantum Information Project, 4-1-8 Honmachi, Kawaguchi, Saitama 331-0012, Japan}
\address{$^3$ ERATO Nuclear Spin Electronics
Project, Aramaki, Aza Aoba,
Sendai 980-0845, Japan}
\ead{tashima@qi.mp.es.osaka-u.ac.jp,
ozdemir@qi.mp.es.osaka-u.ac.jp}

\begin{abstract}
We propose a simple probabilistic optical gate to expand polarization
 entangled W states. The gate uses one polarization-dependent
 beamsplitter and a horizontally polarized single photon as an
 ancilla. The gate post-selectively expands $N$-photon W states to
 $(N+1)$-photon W states. A feasibility analysis considering the
 realistic experimental conditions show that the scheme is within the
 reach of the current quantum optical technologies.
\end{abstract}
\maketitle
\section{Introduction}
Entanglement is the most important resource to realize quantum
information processing tasks which surpass the efficiency of their
classical correspondences as well as provide solutions to problems which
are intractable with the classical resources \cite {s1} - \cite{s3}. The
simplest form of this resource is the one between two parties, so called
bipartite entanglement. Thanks to the theoretical and experimental
efforts within the past two decades, we now have a clear understanding
of the structure and characteristics of bipartite entangled states. It
is well known that any bipartite state can be prepared from a maximally entangled
bipartite state by local operation and classical communication
(LOCC). It is generally accepted that the structure and dynamics of
entanglement become more complex as the number of parties sharing
entanglement increases. This sets a challenge in the theoretical and
experimental studies of multipartite entanglement. Among many
interesting features of multipartite entangled states, the widely-known
one is the presence of inequivalent classes such as
Greenberger-Horne-Zeilinger (GHZ), W and cluster states which cannot be
converted into each other by stochastic local operations and classical
communication (SLOCC) \cite {s7}. Experimental preparation and
characterization of multipartite entangled states thus is not only
essential for a better understanding of the quantum mechanics but also
for the realization of state-specific information processing tasks. For
example, W state has been shown to be the only pure state to exactly solve the
problem of leader election in anonymous quantum networks  whereas
GHZ-state has been shown to be the only pure state to achieve consensus
in distributed networks where no classical post-processing is allowed
\cite {s8}.

The photonic $N$-particle W state is represented by
$\ket{{\rm W}_N}=\ket{(N-1)_{\rm H},1_{\rm V}}/\sqrt{N}$ with
$\ket{(N-k)_{\rm H},k_{\rm V}}$ being the superposition of all
possible permutations of $N-k$ photons with horizontal ($\rm H$)
polarization and $k$ photons with vertical ($\rm V$) polarization, e.g.,
$\ket{{\rm W}_3}=\ket{2_{\rm H},1_{\rm V}}/\sqrt{3}=[\ket {1_{\rm
H}1_{\rm H}1_{\rm V}}+\ket {1_{\rm H}1_{\rm V}1_{\rm H}}+\ket {1_{\rm
V}1_{\rm H}1_{\rm H}}]/\sqrt 3$. W state has the peculiar property that every photon pair has the optimal
amount of pairwise entanglement \cite {s9,s10,s11}. Such an
entanglement structure forms a web-like system where every qubit
has bonds with every other qubit, and the bipartite entanglement
survives even if all the other ($N-2$)-qubits are discarded. In recent
years, there has been a number of theoretical proposals for the
use of W states in multiparty protocols such as QKD \cite {s12},
leader election in anonymous quantum networks \cite {s8} and
teleportation, and for the preparation of W-states in different
systems including optics \cite
{s17.0} - \cite {s18}, ion traps \cite
{s19} and NMR \cite
{s21}. Some of these
proposals have been realized in experiments \cite
{s3,s19}, \cite {s14} - \cite {s17}.
\begin{figure}[h]
\begin{center}
\includegraphics[scale=1]{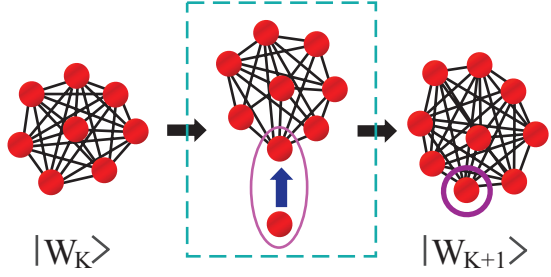}
\caption{Schematic description of local expansion of a W state
using a gate with two inputs, one of which is an ancillary state
and the other is provided from the W state.\label{fig:1s}}
\end{center}
\end{figure}

In a recent study, we introduced an elementary optical gate to
expand a state $\ket{{\rm W}_N}$ to a state $\ket{{\rm
W}_{N+2}}$ by local manipulation on a single site without
accessing all the qubits of the initial W state \cite {s18}. While
such a local expansion was known for GHZ and cluster states, it
was a challenge for W states as (i) the marginal states of the
remaining untouched $N-1$ qubits is different for $\ket{{\rm
W}_N}$ and $\ket{{\rm W}_{N+1}}$ implying that the expansion
process cannot be achieved unitarily, and (ii) the added new qubit
should form pairwise entanglement with each of the untouched $N-1$
qubits of the original W state [see Fig. \ref{fig:1s}]. The gate proposed in Ref. \cite {s18} consists of two 50:50
beamsplitters and a half wave plate (HWP), and expands any W state
by two qubits as the ancillary state used in the gate is an
H-polarized two-photon Fock state. In this paper, on the other
hand, we propose a simpler probabilistic optical gate, which is
based on post-selection, to expand a state $\ket{{\rm W}_N}$ by
only one qubit. This new gate, which is shown in Fig.
\ref{fig:2s}, is composed of a polarization-dependent
beamsplitter (PDBS), an H-polarized single photon as an ancilla and
a HWP for phase compensation. The gate operates as a one-input
two-output gate which can expand the state $\ket{{\rm W}_N}$ to
the state $\ket{{\rm W}_{N+1}}$.

This paper is organized as follows: In Sec. 2, we describe the
principles of the gate operation. Sec. 3 includes a discussion of
how this basic gate structure can be used to expand any
polarization entangled W state. In Sec. 4, we give a scheme for
the experimental realization of this gate and carry out a
feasibility analysis under realistic conditions. Finally, in Sec.
5, we give a brief summary and conclusions.

\section{Gate operation for expanding W state}
\begin{figure}[htbp]
\begin{center}
\includegraphics[scale=1]{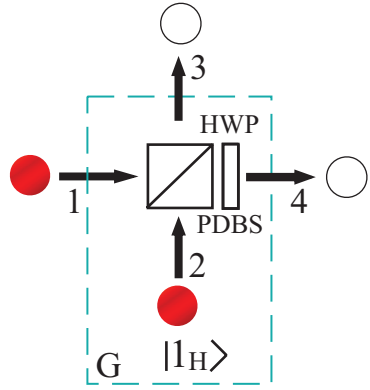}
\caption{The optical gate proposed for the local expansion of W states. The gate uses a
 polarization-dependent beamsplitter (PDBS), a half wave plate (HWP) and
 a horizontally (H) polarized single-photon as an ancilla in input mode
 2. The photon in the input mode 1 comes from the W state to be expanded. \label{fig:2s}}
\end{center}
\end{figure}

The details of the proposed gate are shown in Fig. \ref{fig:2s}.
The key component in this gate is the PDBS whose
reflection and transmission coefficients depend on the polarization of
the input light. The action of a PDBS for H-polarized photons and
V-polarized photons
can be written as
\begin{eqnarray}
\hat{a}^{\dag}_{1{\rm
H}}=\sqrt {1-\mu}~\hat{a}^{\dag}_{3{\rm H}}-\sqrt
{\mu}~\hat{a}^{\dag}_{4{\rm H}}, ~~~\hat{a}^{\dag}_{2{\rm
H}}=\sqrt {\mu}~\hat{a}^{\dag}_{3{\rm H}}+\sqrt
{1-\mu}~\hat{a}^{\dag}_{4{\rm H}},\label{eq:1.1}\end{eqnarray}
and
\begin{eqnarray}\hat{a}^{\dag}_{1{\rm
V}}=\sqrt {1-\nu}~\hat{a}^{\dag}_{3{\rm V}}-\sqrt
{\nu}~\hat{a}^{\dag}_{4{\rm V}}, ~~~\hat{a}^{\dag}_{2{\rm
V}}=\sqrt {\nu}~\hat{a}^{\dag}_{3{\rm V}}+\sqrt
{1-\nu}~\hat{a}^{\dag}_{4{\rm V}}\label{eq:1}
\end{eqnarray}
where $\hat{a}^{\dag}_{j{\rm H}}$ ($\hat{a}^{\dag}_{j{\rm V}}$) denotes the creation operator of
H (V)-polarized photon in the $j$-th mode of PDBS, and $\mu$ ($\nu$) is the
transmission coefficient for H (V)-polarization. The gate uses an H-polarized photon as the ancilla
in mode 2, and a photon in mode 1 with an arbitrary polarization
as the input. The successful operation of the gate is signalled by
a coincidence detection which occurs when there is one photon in
each of the output modes 3 and 4. In order to understand the
working principle of this gate for W-state preparation and
expansion, it is enough to consider its action on two possible
cases: $\ket {1_{\rm H}}_1\ket {1_{\rm H}}_2
=\hat{a}^{\dag}_{1{\rm H}}\hat{a}^{\dag}_{2{\rm H}}\ket{vac}_{12}$
and $\ket {1_{\rm V}}_1\ket {1_{\rm H}}_2 =\hat{a}^{\dag}_{1{\rm
V}}\hat{a}^{\dag}_{2{\rm H}}\ket{vac}_{12}$, where $\ket{vac}$ stands for the vacuum state. Using the relations
given in Eqs. (\ref{eq:1.1}) and (\ref{eq:1}) for the PDBS, we find that these input
states are transformed into
\begin{eqnarray}
\hspace{-25mm}\ket {1_{\rm H}}_1\ket {1_{\rm
H}}_{2}&\rightarrow&\sqrt {2\mu(1-\mu)}~\ket {2_{\rm H}}_3 \ket
{0}_4\underline{+(1-2\mu)\ket {1_{\rm H}}_3\ket {1_{\rm
H}}_{4}}-\sqrt {2\mu(1-\mu)}~\ket {0}_3\ket {2_{\rm
H}}_4,\label{eq:2}
\end{eqnarray} and
\begin{eqnarray}
\hspace{-25mm}\ket {1_{\rm V}}_1\ket {1_{\rm
H}}_{2}&\rightarrow&\sqrt {\mu(1-\nu)}~\ket {1_{\rm V}1_{\rm
H}}_3\ket {0}_{4}+\underline{\sqrt {(1-\nu)(1-\mu)}~\ket {1_{\rm
V}}_3\ket {1_{\rm H}}_{4}}\nonumber\\&&-\underline{\sqrt
{\mu\nu}~\ket {1_{\rm H}}_3\ket {1_{\rm V}}_{4}}-\sqrt
{\nu(1-\mu)}~\ket {0}_3\ket {1_{\rm V}1_{\rm H}}_{4}.\label{eq:3}
\end{eqnarray}
In the above equations, only the underlined terms lead to
successful gate operation and we will focus only on those terms. It is seen that when the input photon is
in V-polarization, the coincidence detection will postselect the
state $\sqrt {(1-\nu)(1-\mu)}~\ket {1_{\rm V}}_3\ket {1_{\rm
H}}_{4}-\sqrt {\mu\nu}~\ket {1_{\rm H}}_3\ket {1_{\rm V}}_{4}$
which is a Bell state if the PDBS parameters are chosen such that
$\mu+\nu=1$. It means that this gate works as an ``entangling gate''. The probability of this event is then $2\mu\nu$.

Next, we see that if the input photon is from a Bell state
$\ket{\rm W_2}=(\ket {1_{\rm V}}_0\ket {1_{\rm H}}_{1}+\ket
{1_{\rm H}}_0\ket {1_{\rm V}}_{1})/\sqrt{2}$, a triple coincidence
at modes 0, 3 and 4 will postselect the state
\begin{eqnarray}
\frac {1}{\sqrt 2}&&[(1-2\mu)\ket {1_{\rm V}}_0\ket {1_{\rm
H}}_3\ket {1_{\rm H}}_{4}+\sqrt {(1-\nu)(1-\mu)}~\ket {1_{\rm
H}}_0\ket {1_{\rm V}}_3\ket {1_{\rm
H}}_{4}\nonumber\\
&&~~-\sqrt {\mu\nu}~\ket {1_{\rm H}}_0\ket {1_{\rm H}}_3\ket
{1_{\rm V}}_{4}]. \label{eq:3a}
\end{eqnarray}
If the weights of the components of this superposition state in
Eq. (\ref{eq:3a}) are made equal, then Eq. (\ref{eq:3a}) will be of
the form $\ket{\rm W_3}$ except a $\pi$-phase shift which can be
compensated using a HWP in mode 4. The equalization of the weights
occurs when
\begin{eqnarray}
1-2\mu=\sqrt {(1-\nu)(1-\mu)}=\sqrt {\mu\nu}.\label{eq:4}
\end{eqnarray}
Second equality in Eq. \ref{eq:4} imposes the condition $\mu+\nu=1$
which is the same condition obtained above for Bell state
preparation. Solving the remaining equalities under the condition
$\mu+\nu=1$, we find that one should choose $\mu=(5-\sqrt 5)/10$ and
$\nu=(5+\sqrt 5)/10$. Inserting these values of $\mu$ and $\nu$ into
Eqs. (\ref{eq:2}) - (\ref{eq:3}),
and imposing the coincidence detection, we find that the
successful gate operation is characterized by the following
transformations
\begin{eqnarray}
\ket {1_{\rm H}}_1\ket {1_{\rm H}}_{2}&\rightarrow&\frac
{1}{\sqrt 5}\ket {1_{\rm H}}_3\ket {1_{\rm
H}}_{4},\nonumber\\\ket {1_{\rm V}}_1\ket {1_{\rm
H}}_{2}&\rightarrow&\frac {1}{\sqrt 5}\ket {1_{\rm V}}_3\ket
{1_{\rm H}}_{4}+\frac {1}{\sqrt 5}\ket {1_{\rm H}}_3\ket {1_{\rm
V}}_{4},\label{eq:8}
\end{eqnarray} where we have included the effect of the HWP in
mode 4. Putting all together, we conclude that this gate can
prepare the Bell state $\ket{\rm W_2}$ with a probability of $2/5$
starting with a V-polarized photon in mode 1, and the $\ket{\rm
W_3}$ state with a probability of $3/10$ starting with the Bell
state $\ket{\rm W_2}$ in modes 0 and 1. This success probability
for $\ket{\rm W_3}$ state preparation is a significant improvement
over other linear optics schemes existing in the literature. Among
the already proposed schemes, the one in Ref. \cite{s18} has
the highest success probability given as $3/16$ which is less than
that of the present scheme.

\section{Expansion of polarization entangled W states}
Here, we show that the same gate can be used to prepare and
expand arbitrary W states. In the following, we will
represent an $N$-partite W state $\ket{{\rm W}_N}$ as $\ket{{\rm
W}_N}=\left[\ket{(N-2)_{\rm H},1_{\rm V}}_{\widetilde{1}}\otimes
\ket{1_{\rm H}}_1+\ket{(N-1)_{\rm H},0_{\rm
V}}_{\widetilde{1}}\otimes\ket{1_{\rm V}}_1\right]/\sqrt{N}$ where
the subscript $1$ denotes the spatial mode of the photon that is
input to the gate and $\widetilde{1}$ denotes the remaining $N-1$
modes of $\ket{{\rm W}_N}$. Using this notation, the
transformation in Eq. (\ref{eq:8}) can be represented as $\ket
{1_{\rm H}}_1\ket {1_{\rm H}}_{2}\rightarrow\sqrt{1/5}\ket
{2_{\rm H},0_{\rm V}}$ and $\ket {1_{\rm V}}_1\ket {1_{\rm
H}}_{2}\rightarrow\sqrt{1/5}\ket {1_{\rm H},1_{\rm V}}$. Thus, we
find that upon the selection of the successful events, the action
of the gate is given as
\begin{eqnarray}
&&\ket{(N-2)_{\rm H},1_{\rm V}}_{\widetilde{1}}\ket{1_{\rm
H}}_1\rightarrow \frac{1}{\sqrt{5}}\ket{(N-2)_{\rm H},1_{\rm
V}}_{\widetilde{1}}\otimes \ket {2_{\rm H},0_{\rm
V}},\nonumber\\
&&\ket{(N-1)_{\rm H},0_{\rm V}}_{\widetilde{1}}\ket{1_{\rm
V}}_1\rightarrow\frac{1}{\sqrt{5}}\ket{(N-1)_{\rm H},0_{\rm
V}}_{\widetilde{1}}\otimes\ket {1_{\rm H},1_{\rm V}}.\label{eq:8a}
\end{eqnarray} Using these relations, it is straightforward to
show that the successful gate operations performs the following
transformation on an initial $\ket{{\rm W}_N}$:
\begin{eqnarray}
\hspace{-15mm}\ket{{\rm W}_N}&\rightarrow&
\frac{1}{\sqrt{5N}}\left[\ket{(N-2)_{\rm H},1_{\rm
V}}_{\widetilde{1}}\otimes \ket {2_{\rm H},0_{\rm
V}}+\ket{(N-1)_{\rm H},0_{\rm V}}_{\widetilde{1}}\otimes\ket
{1_{\rm H},1_{\rm
V}}\right]\nonumber\\\vspace{-10mm}&=&\sqrt{\frac{N+1}{5N}}\ket{{\rm
W}_{N+1}}. \label{eq:8b}
\end{eqnarray} Thus we conclude that the gate expands a given W-state $\ket{{\rm
W}_N}$ to $\ket{{\rm W}_{N+1}}$  by one photon with a success
probability of $(N+1)/5N$. The success probability will approach the constant $1/5$ when $N$ becomes very large. This analysis
shows clearly that the proposed gate can be used in two different
ways: (i) a given arbitrary-size W-state $\ket{{\rm W}_N}$ can be
expanded by one at each successful operation of the gate which
takes place with the probability $(N+1)/5N$, e.g., a
probability of $4/15$ for the expansion of $\ket{{\rm
W}_3}$ to $\ket{{\rm W}_4}$ and (ii) starting from a V-polarized input
photon, an arbitrary-size W-state can be prepared by cascade
application of the gate. For example, cascading $k$ of this gate will
prepare the state $\ket{{\rm W}_{k+1}}$ with a probability of
$(k+1)5^{-k}$.

\section{Practical considerations for an experimental implementation}
In this section, we introduce an experimental scheme for the
implementation of this gate to expand the Bell state $\ket{\rm
W_2}$ to $\ket{\rm W_3}$, and discuss the effects of realistic
conditions on the performance of the gate. We will focus on the
effects of imperfections in (a) the preparation of the $\ket{\rm
W_2}$ and the ancillary state, $\ket{1_{\rm H}}$, (b) the
detection of the successful events, and (c) the deviations of the
parameters of PDBS from its optimal values.

\subsection {Basic scheme}
\begin{figure}[b]
\begin{center}
\includegraphics[scale=1]{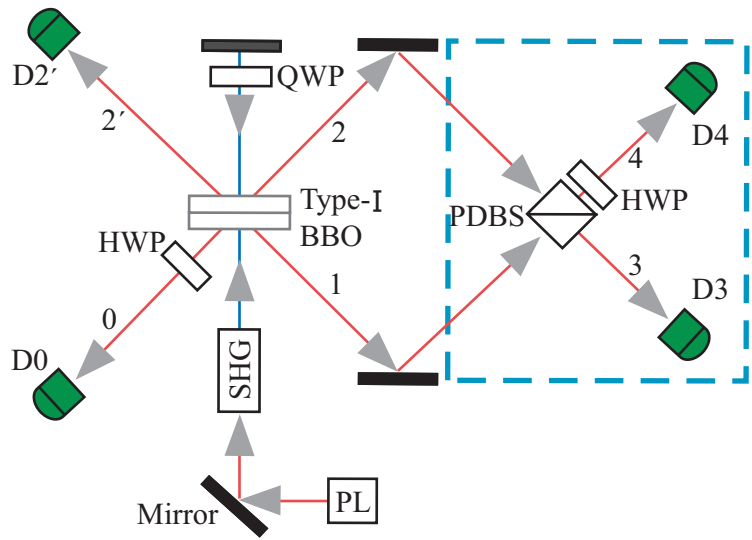}
\caption{Experimental setup for realizing the proposed gate. PL, pulsed laser; SHG, second
harmonic generator;  Type I $\rm BBO$, phase
 matched $\beta$-barium borate crystal for spontaneous
 parametric down conversion (SPDC); PDBS, polarization dependent
 beamsplitters; $\rm D_{j}$, photondetectors; QWP,
 quarter-wave plate; HWP, half-wave plate.\label{fig:3s}}
\end{center}
\end{figure}
We propose the scheme given in Fig. \ref{fig:3s} for the practical
implementation of the proposed gate. In this scheme, the output of
a pulsed laser (PL) with angular frequency $\omega_0$ in the visible
range of the spectrum is frequency doubled in a nonlinear crystal
to produce pulses of ultraviolet (UV) light of angular frequency
$2\omega_0$. These UV pulses are then used to pump twice in
forward and backward directions a pair of nonlinear crystals,
which are stacked together such that their optical axes are
orthogonal to each other \cite {s22.1}. The crystals are for Type-I spontaneous
parametric down conversion (SPDC) to produce photon pairs in two
modes (idler and signal) with the same polarization and at half
the frequency of the pump beam. In the forward pumping direction,
the polarization of the UV beam is set to vertical so that an
H-polarized photon pair in modes 2 and $2^\prime $ are generated from which
the required ancillary state $\ket{1_{\rm H}}$ in mode 2 can be
prepared. The remaining (non-down-converted) portion of the UV
beam first passes through a quarter wave plate (QWP) which changes
its polarization into an ellipsoidal polarization. A mirror placed after the
QWP back-reflects this beam and sends it through the QWP again
which further changes the polarization of the beam into diagonal
polarization. This diagonally
polarized beam pumps the crystals in the backward direction
creating the entangled photon pair $(\ket {1_{\rm H}}_0\ket
{1_{\rm H}}_{1}+\ket {1_{\rm V}}_0\ket {1_{\rm V}}_{1})/\sqrt{2}$.
Changing the polarization of the photon in the mode 0 (idler)
of the SPDC output 
 will prepare the $\ket{\rm W_2}$ in
the spatial modes 0 and 1. Then the ancillary photon in mode 2 and
the photon in mode 1 of $\ket{\rm W_2}$ are mixed at the PDBS. The
successful events are selected by a four-fold coincidence
detection by ON/OFF detectors placed at the
modes 0, $2^\prime$, 3 and 4 as seen in Fig. \ref{fig:3s}.

\subsection {Effects of SPDC and imperfect detection}
Imperfections in the photon detectors affect the gate in two ways:
(i) Recording some of the successful events as the failure due
to non-unit quantum efficiency, and (ii) reporting some of the
failures as the successful due to dark counts and/or due to the
fact that detectors cannot resolve the photon number. In the
following, without loss of generality, we neglect the errors due
to dark counts. This is acceptable as the dark counts rates of
current detectors are very low \cite {s22}. Moreover, the
requirement of four-fold coincidence detection in our scheme
significantly reduces the probability of false events due to dark
counts. Neglecting the dark counts, the positive operator valued measure (POVM) elements for ON/OFF photondetectors become
\begin{eqnarray}
\Pi_{0} = \sum^\infty_{m=0} (1-\eta)^m\ket {m}\bra
{m},\\
\Pi_{1} =1-\Pi_{0}= \sum^\infty_{m=1} [1-(1-\eta)^m]\ket {m}\bra
{m},\label{eq:81e}
\end{eqnarray}
where $\Pi_{0}$ and $\Pi_{1}$ are, respectively, elements
for no click (OFF) and for a click (ON) \cite {s23}. Returning back to our gate,
we see that if there is only one photon in each of the modes 1 and
2, then the success probability of having one photon in each of
modes 3 and 4 becomes $3\eta^4/10$. Note that the error due to
(ii) occurs when there are more than one photon in either or both
of the modes 3 and 4. This takes place when either or both of the
backward and forward SPDC processes prepare two or more photon
pairs. In practical settings, SPDC suffers from the non-deterministic
nature of the process: The output of the SPDC contains vacuum with
high probability and the probability of a photon pair generation
is low. Moreover, although the probability is much lower, there are
cases when multiple pairs of photons are generated. The generated state in the forward direction becomes
\begin{eqnarray}
\ket{\Psi}_{22^\prime }=\sqrt{g}(\ket{vac}_{22^\prime }+\gamma
e^{i\phi_{p}}\ket{1_{\rm H}}_{2}\ket{1_{\rm H}}_{2^\prime}+\gamma^2
e^{2i\phi_{p}}\ket{2_{\rm H}}_{2}\ket{2_{\rm
H}}_{2^\prime}\ldots),\label{eq:81b}
\end{eqnarray}where $g = 1-\gamma^2$ and $\gamma
e^{i\phi_p}$ is proportional to the complex amplitude of the pump
field. Assuming that the
losses in the forward and backward pumping are negligible, the state in the backward direction can be
written as 
\begin{eqnarray}
\ket{\Psi}_{01}=\sqrt{g_1}(\ket{vac}_{01}+\gamma
e^{i\phi_p}\ket{{\rm W_2}}_{01}+\frac {1}{2}\gamma^{2}e^{2i\phi_{p}} \ket{\Lambda}_{01}\ldots),\label{eq:81a}
\end{eqnarray} 
 where $\ket{\Lambda}_{01}=\ket{1_{\rm H}1_{\rm V}}_{0}\ket{1_{\rm
H}1_{\rm V}}_{1}+\ket{2_{\rm V}}_{0}\ket{2_{\rm
H}}_{1}+\ket{2_{\rm H}}_{0}\ket{2_{\rm V}}_{1}$ is unnormalized and
$g_1=(1-\gamma^2/2)^2$. Combining the above expressions, we find that
four-fold coincidence detection postselects the state,
\begin{eqnarray}
\ket{\Psi}_{0122^\prime}&=&\sqrt{gg_1}[~\gamma^2 e^{2i\phi_{p}}\ket{\rm W_2}_{01}\ket {1_{\rm H}}_2\ket {1_{\rm H}}_{2^\prime}\nonumber\\&&+\gamma^3 e^{3i\phi_{p}}(\ket{\rm W_2}_{01}\ket
{2_{\rm H}}_2\ket {2_{\rm H}}_{2^\prime}+\frac {1}{2}\ket{\Lambda}_{01}\ket {1_{\rm
H}}_2\ket {1_{\rm H}}_{2^\prime})]+\mathcal{O}(\gamma^{4})\nonumber\\&=&\sqrt{gg_1}[~\frac {1}{\sqrt 2}\gamma^2 e^{2i\phi_{p}}(\hat{a}^{\dag}_{0{\rm V}}\hat{a}^{\dag}_{1{\rm H}}+\hat{a}^{\dag}_{0{\rm H}}\hat{a}^{\dag}_{1{\rm V}})\hat{a}^{\dag}_{2{\rm H}}\hat{a}^{\dag}_{2^\prime{\rm H}}\nonumber\\&&+\frac {1}{2\sqrt 2}\gamma^3 e^{3i\phi_{p}}(\hat{a}^{\dag}_{0{\rm V}}\hat{a}^{\dag}_{1{\rm H}}+\hat{a}^{\dag}_{0{\rm H}}\hat{a}^{\dag}_{1{\rm V}})(\hat{a}^{\dag}_{2{\rm H}})^2(\hat{a}^{\dag}_{2^\prime{\rm H}})^2\nonumber\\&&+\frac {1}{4}\gamma^3 e^{3i\phi_{p}}\{2\hat{a}^{\dag}_{0{\rm H}}\hat{a}^{\dag}_{0{\rm V}}\hat{a}^{\dag}_{1{\rm H}}\hat{a}^{\dag}_{1{\rm V}}+(\hat{a}^{\dag}_{0{\rm V}})^2(\hat{a}^{\dag}_{1{\rm H}})^2\nonumber\\&&+(\hat{a}^{\dag}_{0{\rm H}})^2(\hat{a}^{\dag}_{1{\rm V}})^2\}\hat{a}^{\dag}_{2{\rm H}}\hat{a}^{\dag}_{2^\prime{\rm H}}]\ket {vac}_{0122^\prime}+\mathcal{O}(\gamma^{4}),\label{eq:81c}
\end{eqnarray}
 where we have focused on the terms up to $\gamma^3$ by considering that in practice, $\gamma^2\sim
\mathcal{O}(10^{-4})$ is very small. The PDBS transforms modes 1 and 2
of $\ket{\Psi}_{0122^\prime}$ according to the relations given in
Eqs. (\ref{eq:1.1}) and (\ref{eq:1}). Let
$\ket{\Psi^\prime}_{0342^\prime}$ be the state after the
transformation. Using POVM given in Eq. (\ref{eq:81e}), the four-fold coincidence
detection probability $p_c$ can be calculated as 
\begin{eqnarray}
p_c &=&{_{0342^\prime}\bra {\Psi^\prime}}{\Pi}^{0}_{1}{\Pi}^{3}_{1}{\Pi}^{4}_{1}{\Pi}^{2^\prime}_{1}\ket{\Psi^\prime}_{0342^\prime}\nonumber\\&=&\underbrace {\frac {1}{2}gg_1\gamma^4\eta^4[2\mu(\mu-1)+1]}_{p_t}\nonumber\\&&\underbrace {+\frac {1}{2}gg_1\gamma^6\eta^4(2-\eta)^2+\frac {1}{2}gg_1\gamma^6\eta^4(2-\eta)^2[\mu(\mu-1)+1]+\mathcal{O}(\gamma^{8})}_{p_f},\label{eq:81cc}
\end{eqnarray} 
where ${\Pi}^{j}_{1}$ is the POVM for ``click'' events at the detection in
mode $j$. $p_t$ and $p_f$ respectively corresponds to probability of
true and false coincidences. We see in Eq. (15) two contributions, $p_t$
and $p_f$. The true coincidences ($p_t$) are due to the $\ket{\rm
W_2}_{01}\ket {1_{\rm H}}_2\ket {1_{\rm H}}_{2^\prime}$ term in
$\ket{\Psi}_{0122^\prime}$ and the false coincidences ($p_f$) originates
from multiple pairs of photons. Plugging the
value $\mu=(5-\sqrt 5)/10$ in these terms, we find that
the ratio of the true coincidences to the total coincidence
events becomes
\begin{eqnarray}
p=\frac{p_t}{p_t+p_f}= 1-3\gamma^2(\eta-2)^2+\mathcal{O}(\gamma^{4}).\label{eq:819a}
\end{eqnarray}
It is clearly seen that almost all the four-fold coincidence detections
are true coincidences within the range of realistic values of $\eta$ and
$\gamma$.

\subsection{Effect of deviation in the PDBS parameter} 
\begin{figure}[pbh]
\begin{center}
\vspace{3mm}
\includegraphics[scale=1]{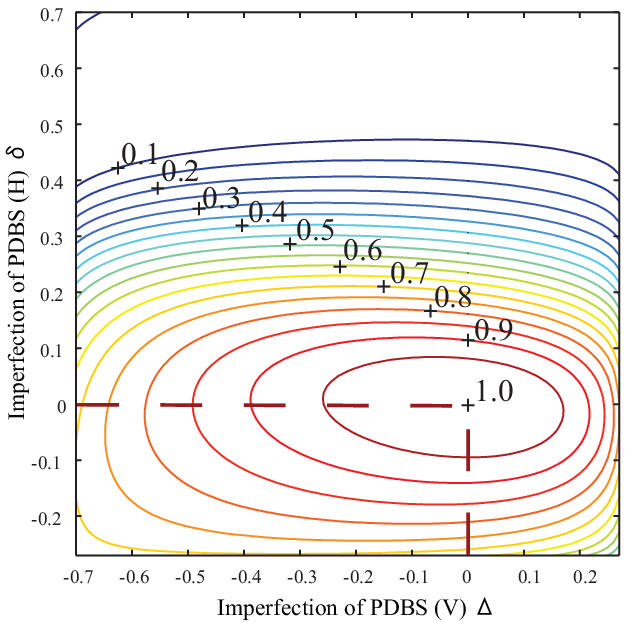}\hspace{3mm}\includegraphics[scale=1]{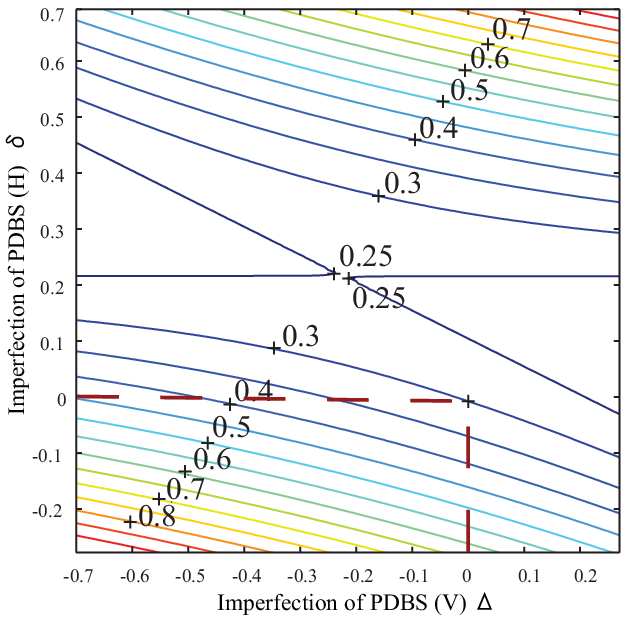}
\caption{Constant fidelity (left) and constant probability (right)
contour plots as a function of $\delta$ and $\Delta$, which are deviations of the parameters $\mu$ and $\nu$ of PDBS from their ideal values, respectively.
\label{fig:5s}}
\end{center}
\end{figure}

In this section, we consider the effect of deviations in the parameters of
PDBS from its ideal values of $\mu=(5-\sqrt 5)/10$ and $\nu=(5+\sqrt 5)/10$ on the probability
and the fidelity of expanding $\ket {\rm W_2}$ into $\ket {\rm
W_3}$. Let us assume that the reflection coefficients of PDBS for
H- and V-polarized photons are deviated from the ideal values by $\delta$ and $\Delta$, respectively. Then the action of the
imperfect PDBS on H-polarized light and V-polarized light becomes
\begin{eqnarray}
\hspace{-18mm}\hat{a}^{\dag}_{1{\rm H}}=\sqrt
{1-\mu-\delta}~\hat{a}^{\dag}_{3{\rm H}}-\sqrt
{\mu+\delta}~\hat{a}^{\dag}_{4{\rm H}}, ~~ \hat{a}^{\dag}_{2{\rm
H}}=\sqrt {\mu+\delta}~\hat{a}^{\dag}_{3{\rm H}}+\sqrt
{1-\mu-\delta}~\hat{a}^{\dag}_{4{\rm H}},
\end{eqnarray}and
\begin{eqnarray}
\hspace{-18mm}\hat{a}^{\dag}_{1{\rm V}}=\sqrt
{1-\nu-\Delta}~\hat{a}^{\dag}_{3{\rm V}}-\sqrt
{\nu+\Delta}~\hat{a}^{\dag}_{4{\rm V}}, ~~ \hat{a}^{\dag}_{2{\rm
V}}=\sqrt {\nu+\Delta}~\hat{a}^{\dag}_{3{\rm V}}+\sqrt
{1-\nu-\Delta}~\hat{a}^{\dag}_{4{\rm V}},
\end{eqnarray}
where $-\mu\leq\delta\leq\nu$ and $-\nu\leq\Delta\leq\mu$. Using these expressions, we calculated
the probability of coincidence detection and the fidelity of the
output state to the desired one. We omit the analytic expressions since they are
rather lengthy and complicated. Instead, we
depict the constant fidelity and constant probability contours as
a function of $\delta$ and $\Delta$ in Fig. \ref{fig:5s}. We see
that the effect of $\delta$ on the fidelity
is much larger than that of $\Delta$. We can thus tolerate larger
deviations from the ideal value for $\Delta$.
\section{Conclusion}
In this paper, we have proposed a simple probabilistic optical
gate for expanding polarization entangled W states and analyzed
its feasibility taking into account the imperfections encountered
in practice. The proposed gate is based on post-selection process
to expand $\ket{{\rm W}_{N}}$ by one qubit into $\ket{{\rm
W}_{N+1}}$ by locally acting on one of its qubits. A remarkable
feature of this gate is that starting with a Bell state, it can
prepare tripartite entangled W-state with a success probability
of $3/10$ which is the highest among all the proposed schemes so
far. Moreover, the gate does not need stabilization of optical
paths and does not employ sub-wavelength adjustments. Our
feasibility analysis shows that the proposed gate can be
implemented by the current experimental technologies. Thus this gate will
provide a simple and useful tool to probe interesting features of
multipartite W states.
\section*{Acknoledgements}
This work was supported by the 21st Century GCOE Program by the Japan Society
for the Promotion of Science, the Japan Society for the Promotion of
Science (JSPS) Grant-in-Aid for Scientific
Research (C) 20540389 and the Ministry of Education, Culture, Sports,
Science and Technology-Japan (MEXT) Grant-in-Aid for Young
Scientists (B) 20740232.

\end{document}